\documentclass[twocolumn]{aastex631}

\shorttitle{PBH dark matter lensing}
\shortauthors{Wen \& Kemball}

\begin{document}

\title{Testing Primordial Black Hole Dark Matter with ALMA Observations of the Gravitational Lens B1422+231}


\author[0000-0003-2812-8607]{Di Wen}
\affil{Department of Astronomy, University of Illinois at Urbana-Champaign, 1002 W. Green Street, Urbana, IL 61801, USA}
\affil{Kapteyn Astronomical Institute, University of Groningen, P.O. Box 800, 9700AV Groningen, The Netherlands}
\author[0000-0001-6233-8347]{Athol J. Kemball}
\affil{Department of Astronomy, University of Illinois at Urbana-Champaign, 1002 W. Green Street, Urbana, IL 61801, USA}

\begin{abstract}

We examine the flux density ratio anomaly in the quadruply-imaged strong gravitational lens, B1422+231, and consider the contribution of $10-10^3M_{\odot}$ primordial black holes (PBHs) as a potential dark matter constituent. We describe the first flux density ratio measurement of B1422+231 in the millimeter-wave band using the Atacama Large Millimeter Array (ALMA). This fills an important multi-wavelength gap in our knowledge of this key lensed system. The flux density of the quasar at 233 GHz is dominated by synchrotron emission and the source size is estimated to be 66.9 pc. The observed flux density ratios at 233 GHz are similar to those measured in radio, mid-infrared and optical bands, which cannot be explained by a simple smooth mass model of the lens galaxy. We examine the probability of the flux density ratio anomaly arising from PBH microlensing using ray tracing simulations. The simulations consider the cases where 10\% and 50\% of dark matter are $10-10^3M_{\odot}$ PBHs with a power law mass function. Our analysis shows that the anomalous flux density ratio for B1422+231 can be explained by a lens model with a significant fraction of dark matter being PBHs. This study demonstrates the potential for new constraints on PBH dark matter using ALMA observations of multiply imaged strong gravitational lenses.

\end{abstract}


\section{Introduction} \label{sec:intro}

The nature of dark matter remains as a puzzle. It is plausible that a fraction of the dark matter in galaxies and their surrounding dark matter halos could come from primordial black holes (PBHs). The idea that a fraction of dark matter could be PBHs in the mass range $10- 10^3 M_{\odot}$ has become more popular \citep[e.g.][]{bird16,carr16} after the gravitational wave detections of merging binary black holes with masses above $10M_{\odot}$ \citep{abbott16,abbott19ligo}. Observational constraints on the fraction of dark matter in a wide range of compact object masses show that PBHs of mass $1-10^3 M_{\odot}$ might be a candidate for dark matter \citep{carr16}. The mass and abundance of PBH dark matter are constrained by microlensing \citep{green16,mediavilla17}, star clusters \citep{brandt16,green16}, dwarf galaxies \citep{koushiappas17,zhu18}, and wide binary stars \citep{quinn09,yoo04}. Recently, there has been strong observational evidence for a 150 $M_{\odot}$ intermediate-mass black hole (IMBH), the remnant of a binary black hole merger \citep{abbott20}. Such IMBHs could also be PBHs. At highest accuracy, constraints on the fraction of dark matter as PBHs have to be calculated based on an extended PBH mass function. PBHs with a unified multi-modal mass function that peaks at around $10^{-5}$, $2$, $30$ and $10^6 M_{\odot}$ could provide the dark matter and satisfy a variety of observational constraints \citep{carr20,carr21a}. A recently proposed method calculates the allowed mass range of primordial black holes as dark matter based on published constraints for the black hole mass function \citep{carr16}. In particular, bounds from microlensing and accretion show $10-10^3 M_{\odot}$ PBHs can constitute a fraction or all of dark matter \citep{carr20,carr21b}. Unfortunately, the mechanism for forming IMBHs, either from massive stars or primordial black holes, is not well understood \citep[see][for review]{miller04}.

Gravitational lensing provides some of the most compelling and direct evidence for dark matter \citep[e.g.][]{clowe06}. General relativity predicts that light rays are deflected by massive objects. The deflection of light due to a gravitational lens galaxy or galaxy cluster can distort the shapes of the source galaxy images in the case of weak gravitational lensing, or produce multiple distorted images of the source galaxy in the case of strong gravitational lensing \citep{schneider92,kochanek06}. Furthermore, different mass distribution in the gravitational lens produce different image positions, shapes and flux densities. This makes gravitational lensing one of the most powerful tools for probing the distribution of dark matter.

Numerical simulations have shown that in the cold dark matter paradigm hierarchical structure formation produces a large population of dark matter substructures (DMSs) within galaxy-sized and cluster-sized dark matter halos \citep{klypin99,moore99}. The DMSs should potentially host dwarf galaxies or satellite galaxies, but the observed abundance of dwarf galaxies does not agree with the predicted abundance of DMSs \citep{bullock10,bullock17}. This `missing satellite problem' has recently been resolved by taking observational selection function into account \citep{nadler21}. The distribution of dark matter in dark matter halos is clumpy because of the presence of DMSs. Their signatures on the lensed images make galaxy-galaxy strong gravitational lenses the best targets for detecting DMSs observationally. Several techniques have been developed for detecting individual DMS \citep{vegetti09,vegetti12,hezaveh13b,macleod13,nierenberg14,hezaveh16a}, measuring the DMS power spectrum \citep{hezaveh16b,cyr19}, and extracting the halo mass function \citep{gilman20,ostdiek22} with strong gravitational lensing. DMSs near a strongly lensed image, either inside the lens galaxy halo or along the line of sight, perturb the smooth gravitational potential of the lens halo and change the magnification and the flux densities of lensed images \citep{dalal02,chen03,metcalf05,wambsganss05,xu12}. This leads to anomalous flux density ratios that cannot be explained by a smooth lens. The presence of DMSs affects the flux density ratios observed in all wave bands.

DMSs may not be the only dark perturbers causing flux density ratio anomalies. A population of PBHs as a fraction of dark matter could be another non-smooth mass component of the lens galaxy. Compact perturbers with masses $10-10^5 M_{\odot}$ are considered IMBHs, which can be of primordial origins. A stellar component on top of smoothly distributed dark matter in the lensing galaxy has been shown to enhance microlensing fluctuations and broadens the magnification probability distribution toward both magnification and demagnification \citep{schechter02,schechter04}. The flux density ratio anomalies observed in optical and X-ray bands are usually caused by stellar microlensing. Similarly, microlensing by a population of IMBHs can produce additional magnification on top of a macro-lensed image, producing an uneven magnification distribution. The distortion pattern of an individual IMBH along the line of sight in the lensing galaxy has an angular scale of a few $\mu$as to mas. Even though individual microlens' Einstein radius is much smaller than the source size, the net magnification due to microlensing by a population of stars converges to a constant low magnification for large source sizes, shown by \citet{barvainis02} for 30 lensed quasars at submillimeter bands. A population of IMBHs can in principle produce analogous effects and have a significant contribution to the mm-wave flux density ratio anomaly. IMBHs could also be the answer to the issue raised by \citet{xu15} that DMSs are unlikely to be the full reason for lensed flux density ratio anomalies in the radio band. \citet{schechter04} demonstrated using microlensing simulations that the magnification probability distribution depends on the mass spectrum of the compact objects. Assuming a fraction of dark matter is PBHs, a strongly lensed quasar could show flux density ratio anomalies caused by quasar microlensing where PBHs act as microlenses. The degree of flux density ratio anomaly observed in a multiply imaged, strong gravitational lens system could place new constraints on PBH dark matter.

In this paper, Section \ref{1422:obs} describes the first flux density ratio measurement in the mm-wave band for the strong gravitational lens, B1422+231. Section \ref{smoothmodel} presents the lens and source models for the ALMA observations. In Section \ref{1422:sim} and \ref{1422:results}, we examine whether PBHs alone can produce the quadruply-lensed flux density ratio anomalies by performing PBH microlensing simulations.

\section{ALMA Observations of B1422+231} \label{1422:obs}
\begin{figure}[t]
\centering
\includegraphics[width=1.0\linewidth]{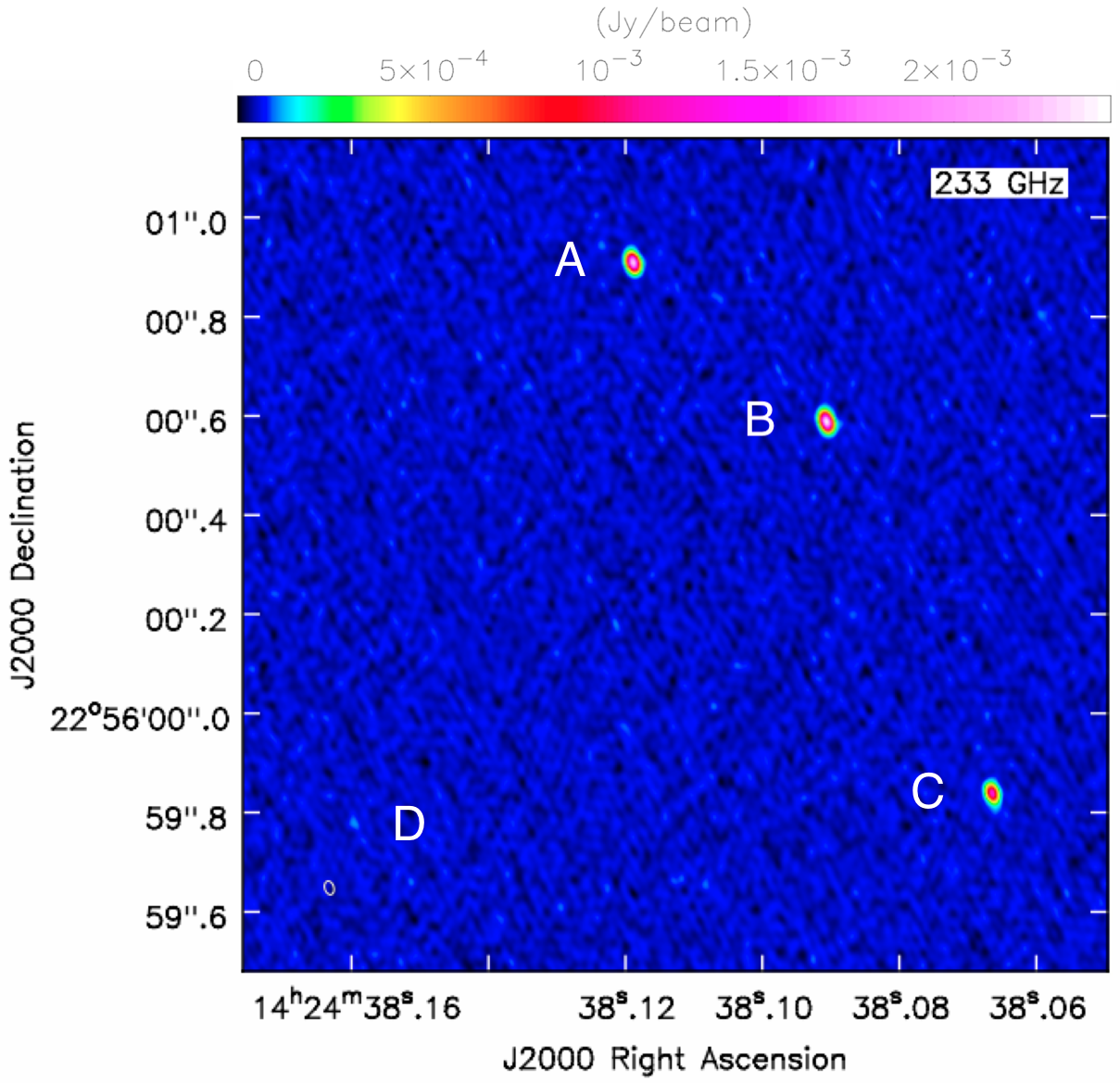}
\caption{Naturally weighted ALMA Band 6 continuum map of B1422+231 at 233 GHz over a 7.5 GHz total bandwidth. The size of the restoring beam is $0.045^{\prime\prime}\times0.022^{\prime\prime}$, shown in the bottom-left corner. The rms noise level is 11 $\mu$Jy beam$^{-1}$.}
\label{imaging}
\end{figure}
B1422+231 is a quadruply-lensed radio-loud quasar that is known for its flux density ratio anomaly. It has previously been observed in radio, infrared, optical and X-ray wave bands. We observed B1422+231 for the first time in the mm-wave band with ALMA. Continuum observations in Band 6 at 233 GHz over a total bandwidth of 7.5 GHz were carried out in two execution blocks on 2019 June 19 and 27 with ALMA using a hybrid configuration of C43-9 and C43-10 (Project ID: 2018.1.00915.S). The ALMA array consisted of 43 long baseline antennas and baseline lengths ranged from 83.1 m to 16.2 km. The calibrators were J1256-0547, J1427+2348 and J1436+2321. The total on-source integration time was 69.02 minutes. The data collected in the first execution block passed quality-assurance stage 0 and 2. The data collected in the second execution block were classified ``SemiPass" in quality-assurance stage 0 due to bad phase transfer from the phase calibrator to the target. The Common Astronomy Software Applications package (CASA; \citep{mcmullin07}) and ALMA pipeline (Version 6.2.1.7; Pipeline Version 2021.2.0.128) were used for calibration and imaging for data from both execution blocks. The measurement set from each execution block was calibrated separately, first with the ALMA pipeline and then phase self-calibrated. The phase self-calibration used a 80s solution interval for the first execution block and a 70s solution interval for the second execution block. The image signal-to-noise (S/N) for the second execution block is 2.3 times lower than that of the first execution block, due to the large noise in phase. Amplitude and phase calibration were performed after combining the two phase self-calibrated measurement sets in order to align the flux scales measured on two observation dates, assuming the source was not variable. To improve sensitivity to the weakest demagnified lensed image, imaging was carried out using natural weighting of the visibilities. Figure \ref{imaging} shows the final image of the self-calibrated combined measurement sets.

The total continuum flux density of B1422+231 at 233 GHz we measure is $6.305\pm0.082$ mJy, summing the flux densities from all four lensed images. This is slightly lower than the measured flux density of $7.5\pm0.8$ mJy at 238 GHz by the Submillimeter Array \citep{keating18}. The missing flux density is likely caused by the fact that our maximum recoverable scale (MRS) is smaller than the extended emission along the lensing arcs near the image triplet (see Figure 3). A MRS of $0.322^{\prime\prime}$ is derived for the hybrid ALMA configuration, based on the $5^{\text{th}}$ percentile baseline length 810.7 meters. The angular sizes of the arcs are expected to be comparable to the image separations ($\sim0.50^{\prime\prime}$ between image A and B and $\sim0.82^{\prime\prime}$ between image B and C). The delivered data were incomplete relative to our proposal request due to proposal scheduling priority. We requested hybrid observations including the more compact C43-6 configuration, but no data was taken with this configuration. Assuming the radio synchrotron spectral index $-1.35$ derived from measurements at 15.0 and 22.5 GHz \citep{tinti05,stacey18}, we expect a flux density of 6.13 mJy at 233 GHz and 5.96 mJy at 238 GHz. Our measured flux density is consistent with these values, suggesting the detected compact emission at in Band 6 is mostly dominated by the synchrotron emission from the quasar's active galactic nucleus (AGN) with almost no thermal dust emission from the dusty torus expected to surround the accretion disk \citep{urry95}.
\begin{figure}[t]\label{fluxratioobs}
\centering
\includegraphics[width=1.0\linewidth]{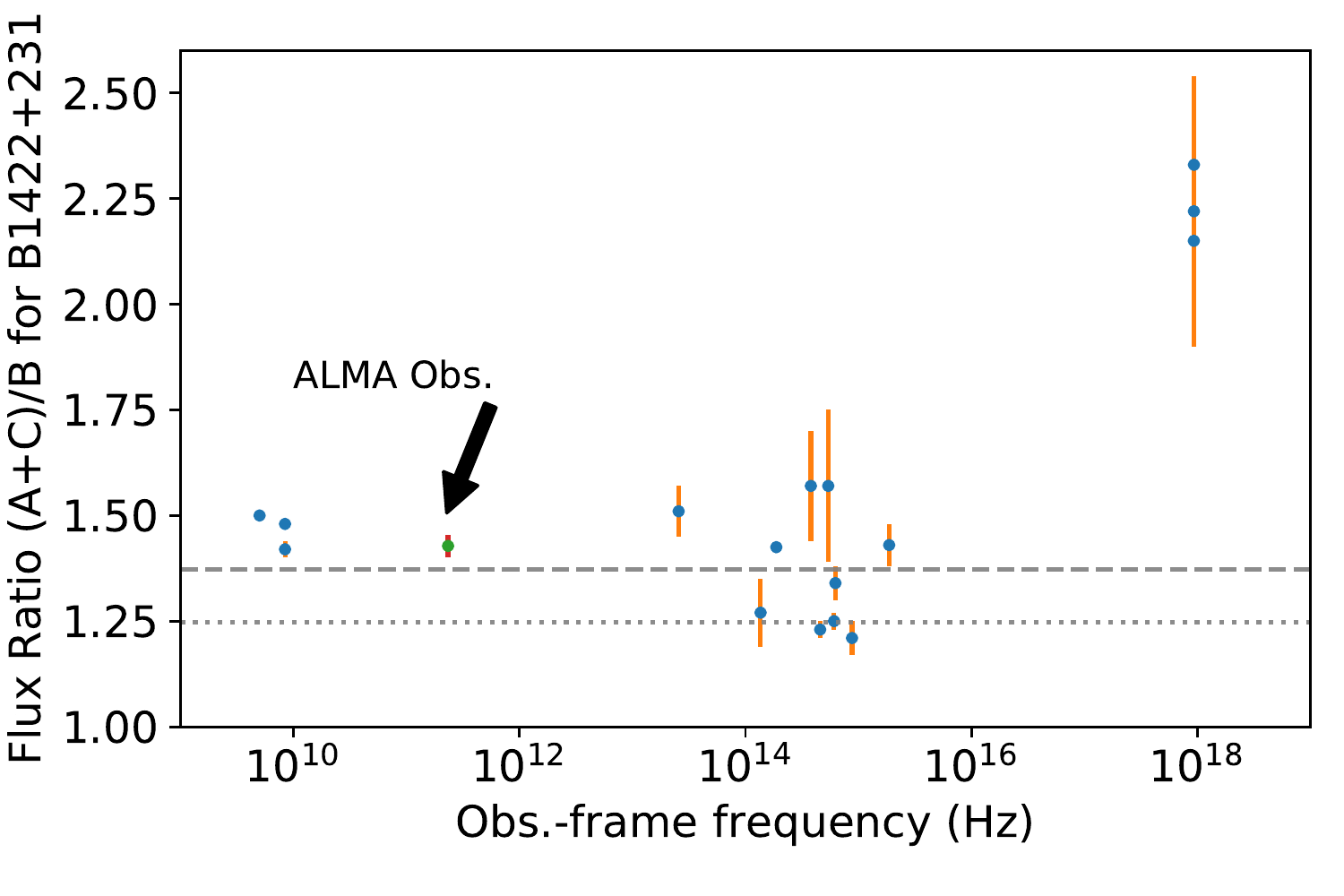}
\caption{Flux density ratio anomalies as a function of observing frequency \citep{patnaik92,lawrence92,impey96,chiba05,sluse12,pooley12}. Our ALMA Band 6 observation labeled with an arrow shows a flux density ratio of $(A+C)/B = 1.434\pm0.017$, consistent with measurements in radio and mid-infrared frequencies. The dotted and dashed horizontal lines show the predictions of the best-fit smooth lens model by \citet{xu15}, (A+C)/B=1.247 and by \citet{schechter14}, (A+C)/B=1.372. Data points with large deviation from the horizontal lines exhibit large flux density ratio anomalies.}
\end{figure}

\begin{deluxetable*}{cccccccccc}[th]\label{components}
\tablecaption{B1422+231 image positions and flux densities measured in ALMA Band 6 continuum observations. The major axis and the minor axis are the FWHM of the best-fit 2-dimensional Gaussian image component sizes after deconvolution from a $0.045^{\prime\prime}\times0.022^{\prime\prime}$ restoring beam with position angle 24.756$^\circ$. The phase center coordinates are J2000 RA 14:24:38.1168, Dec +22:56:00.175.}
\tablehead{
\colhead{Image} & \colhead{Right Ascen.} & \colhead{Declination} & \colhead{$\Delta\alpha\,\rm cos(\delta)$} & \colhead{$\Delta\delta$} & \colhead{Flux Density} & \colhead{Flux Density Ratio} & \colhead{Major Axis} & \colhead{Minor Axis} & \colhead{PA}\\
 & \colhead{(h m s)} & \colhead{($^{\circ}$ $^{\prime}$ $^{\prime\prime}$)} & \colhead{(arcsec)} & \colhead{(arcsec)} & \colhead{(mJy)} &  & \colhead{(mas)} & \colhead{(mas)} & \colhead{(deg)}
 }
\startdata
A & 14:24:38.118 & +22:56:00.890 & +0.017 & +0.715 & 2.338$\pm$0.019 & 0.912$\pm$0.011 & 5.87$\pm$0.97 & 2.62$\pm$1.48 & 35$\pm$14\\
B & 14:24:38.090 & +22:56:00.570 & $-$0.370 & +0.395 & 2.565$\pm$0.023 & 1 & 8.53$\pm$0.70 & 0.76$\pm$1.25 & 23.9$\pm$4.3\\
C & 14:24:38.066 & +22:55:59.823 & $-$0.702 & $-$0.355 & 1.339$\pm$0.022 & 0.522$\pm$0.010 & 7.8$\pm$1.3 & 1.2$\pm$1.3 & 16.5$\pm$8.7\\
D & 14:24:38.158 & +22:55:59.762 & +0.569 & $-$0.415 & 0.063$\pm$0.018 & 0.025$\pm$0.007 & - & - & -
\enddata
\end{deluxetable*}

\subsection{Flux Density Ratio of Image Components} \label{fluxratio}
Four unresolved point sources are detected and labeled as image A, B, C, and D (see Figure \ref{imaging}). Image A, B, and C correspond to the brightest three image components in a cusp configuration of the quadruply-imaged B1422+231. The foreground lens galaxy is not detected, which is expected at this frequency. Image A, B, and C are individually fitted with two-dimensional elliptical Gaussian image components parametrized by peak intensity, peak pixel coordinates, major and minor axes, and position angle using CASA. The best-fit image properties are presented in Table \ref{components}. The image positions in our observation are consistent with those from radio, infrared and optical observations \citep{patnaik92,lawrence92,impey96,nierenberg14}. The apparent alignment of position angles among A, B and C in Figure \ref{imaging} only reflects the position angle of the beam, which has higher angular resolution along the tangential direction of the triplet's arc than the radial direction. The uncertainties of the fitted position angles for the three image components after devolution from the beam (in the last column of Table \ref{components}) are likely underestimated. The error estimates of the integrated flux to image components are calculated during the Gaussian fitting implemented in CASA. The error estimates to the fitting parameters are based on the formulation by \citet{condon97}. The image component positions from Gaussian fitting before self-calibration are the same as those in Table \ref{components}. Image component D has the largest position uncertainties from Gaussian fitting due to its low S/N, 1.3 mas in right ascension and 2.6 mas in declination. Figure \ref{fluxratioobs} shows the flux density ratio of B1422+231 measured across the electromagnetic spectrum. We find that the flux density ratio (A+C)/B for B1422+231 at 233 GHz to be $1.434\pm0.017$, consistent with the values from radio, mid-infrared and most optical observations. The frequency-independent flux density ratios in radio, mm and mid-infrared observations suggest that the emissions at these frequencies likely originate close to the accretion disk of the AGN. Most flux density ratios across all frequencies in Figure \ref{fluxratioobs} deviate significantly from the prediction from the best smooth lens model by \citet{xu15}, which indicates that the mass distribution in the foreground lens galaxy is not smooth. The presence of DMSs is likely the primary contributor to the observed flux density ratio anomalies at low frequencies. The large scatter and deviation from the smooth model prediction in optical and X-ray observations suggests that stellar microlensing likely causes additional flux density ratio anomaly contributions.

\begin{deluxetable*}{cccccccccc}\label{table:glaficfit}
\tablecaption{Best-fit lens model parameters. The first two lines are fitted using \texttt{glafic}. Note that $\Delta x_s$ and $\Delta x_l$ have opposite signs compared to $\Delta\alpha\,\rm cos(\delta)$ in Table \ref{components}. The position angle $\theta_{e}$ is defined in degrees east of north in \texttt{glafic}, where +x direction is west and +y direction is north. The SIE (${\bar S}_{\nu}$) model uses only image component positions as constraints and not flux densities. The SIE ($S_{\nu}$) model uses both image component positions and flux densities as constraints. A SIE ({[O {\scriptsize III}]}) model from \citet{nierenberg14} is included for comparison. }\label{tab:deluxesplit}
\tablehead{
\colhead{Model} & \colhead{$\Delta x_s$} & \colhead{$\Delta y_s$} & \colhead{$\Delta x_l$} & \colhead{$\Delta y_l$} &\colhead{$\theta_{Ein}$} & \colhead{$e$} & \colhead{$\theta_e$} & \colhead{$\gamma_x$} & \colhead{$\theta_{x}$}\\
  & \colhead{(arcsec)} & \colhead{(arcsec)} & \colhead{(arcsec)} & \colhead{(arcsec)} & \colhead{(arcsec)} &  & (deg) &  & (deg)
}
\startdata
SIE (${\bar S}_{\nu}$) & $-$0.071 & $-$0.070 & $-$0.355 & $-$0.264 & 0.7755 & 0.285 & 127.3 & 0.172 & 125.0\\ 
SIE ($S_{\nu}$) & $-$0.066 & $-$0.068 & $-$0.423 & $-$0.310 & 0.7996 & 0.477 & 126.6 & 0.117 & 125.0\\
SIE ({[O {\scriptsize III}]}) & - & - & $-$0.3896 & $-$0.2404 & 0.771 & 0.16 & 123 & 0.22 & 126
\enddata
\end{deluxetable*}

\begin{deluxetable*}{cccccccccc}\label{table:visilensfit}
\tablecaption{Best-fit S{\'e}rsic source model parameters using \texttt{visilens}. The source position angle $\phi_s$ is defined in degrees counter-clockwise from the lens major axis (see Figure 3), where +x direction is west and +y direction is north. The last column lists the relative log-evidence. The SIE ({[O {\scriptsize III}]}) model assumes the lens model in Table \ref{table:glaficfit} \citep{nierenberg14}.}
\tablehead{
\colhead{Lens Model} & \colhead{$\Delta X_s$} & \colhead{$\Delta Y_s$} & \colhead{$F_s$} & \colhead{$a_s$} & \colhead{$n_s$} & \colhead{$b_s/a_s$} & \colhead{$\phi_s$} & \colhead{$\Delta$log$Z$}\\
  & \colhead{(arcsec)} & \colhead{(arcsec)} & \colhead{($\mu$Jy)} & \colhead{(mas)} &  &  & (deg)
}
\startdata
SIE (${\bar S}_{\nu}$) & $0.34436^{+0.00052}_{-0.00073}$ & $0.23053^{+0.00025}_{-0.00037}$ & $0.2963^{+0.0043}_{-0.0038}$ & $9.03^{+0.15}_{-0.12}$ & $0.050^{+0.011}_{-0.015}$ & $0.850^{+0.024}_{-0.026}$ & $105^{+20}_{-10}$ & 0\\
SIE ({[O {\scriptsize III}]}) & $0.36007^{+0.00093}_{-0.00025}$ & $0.24717^{+0.00063}_{-0.00070}$ & $0.97^{+0.77}_{-0.71}$ & $9.07^{+1.10}_{-0.96}$ & $0.106^{+0.057}_{-0.058}$ & $0.166^{+0.066}_{-0.034}$ & $72.88^{+0.56}_{-1.0}$ & $-$2589
\enddata
\end{deluxetable*}

\section{Lens and Source Model of B1422+231}\label{smoothmodel}
The starting point of our mass model for B1422+231 is a singular isothermal ellipsoid (SIE) plus external shear $\gamma_x$ (SIE$+\gamma_x$) model. Optical and near-infrared observations have shown that B1422+231 belongs to a group consisting of five nearby galaxies that provide sufficient external shear to produce the observed image configuration \citep{kundic97}. The source galaxy is at redshift $z_s=3.62$ and the foreground lens galaxy is at redshift $z_l=0.34$ \citep{patnaik92,tonry98}. We modeled the ALMA observation of B1422+231 in three steps: 1) lens modeling with image positions and/or flux densities as constraints assuming a point source; 2) visibility-space parametric source modeling while holding the lens model fixed; and 3) visibility space joint parametric modeling of the lens and source. A Hubble constant $H_0 = 67.7$ km Mpc$^{-1}$ s$^{-1}$ and a matter density parameter $\Omega_M = 0.307$ \citep{planckxiii16} are assumed.

\subsection{Lens Model}
We first used only the image positions in Table \ref{components} as constraints and modeled the lens system with the software package \texttt{glafic} \citep[version 2;][]{oguri10}, which minimizes $\chi^2$ in the source plane. The SIE+$\gamma_x$ model includes seven parameters (columns 4-10 in Table \ref{table:glaficfit}): relative positions of the dark matter halo center to the phase center in RA and DEC directions $\Delta x_l$ and $\Delta y_l$, Einstein radius $\theta_{Ein}$, ellipticity of the lens halo $e$, position angle of the lens dark matter halo $\theta_e$, external shear $\gamma_x$ and position angle of the external shear $\theta_{x}$. The source model is a point source with coordinates $\Delta x_s$ and $\Delta x_y$ relative to the phase center. We show the best-fit lens model with both the image positions and flux densities as constraints. However, because of the known flux density ratio anomaly of this system, the best-fit lens model using only image position as constraints is more reliable than that with flux densities as constraints. Table \ref{table:glaficfit} shows that our model with only position constraints is similar to the published model obtained for [O {\scriptsize III}] images using only image positions as constraints \citep{nierenberg14}.

\begin{figure}[b]\label{modelimage}
\centering
\includegraphics[width=1.0\linewidth]{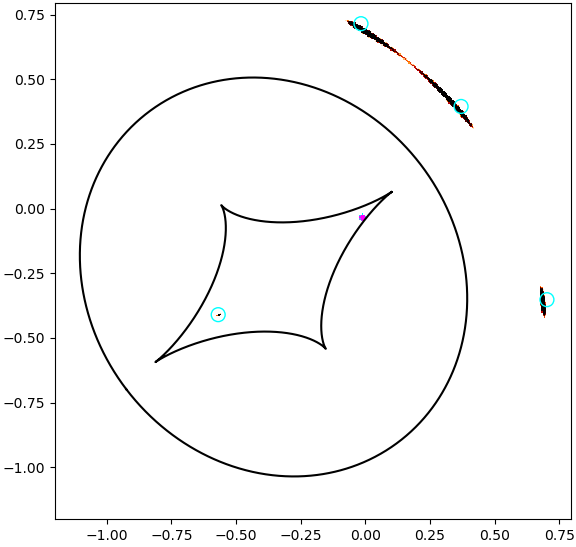}
\caption{The best-fit model image (black-orange) from the SIE (${\bar S}_{\nu}$) lens model and the best-fit S{\'e}rsic source profile (pink). The source-plane caustics are plotted in black solid lines. The observed image positions are labeled with cyan circles.}
\end{figure}

\begin{figure*}[tbh]\label{fig:corner}
\centering
\includegraphics[width=1.0\linewidth]{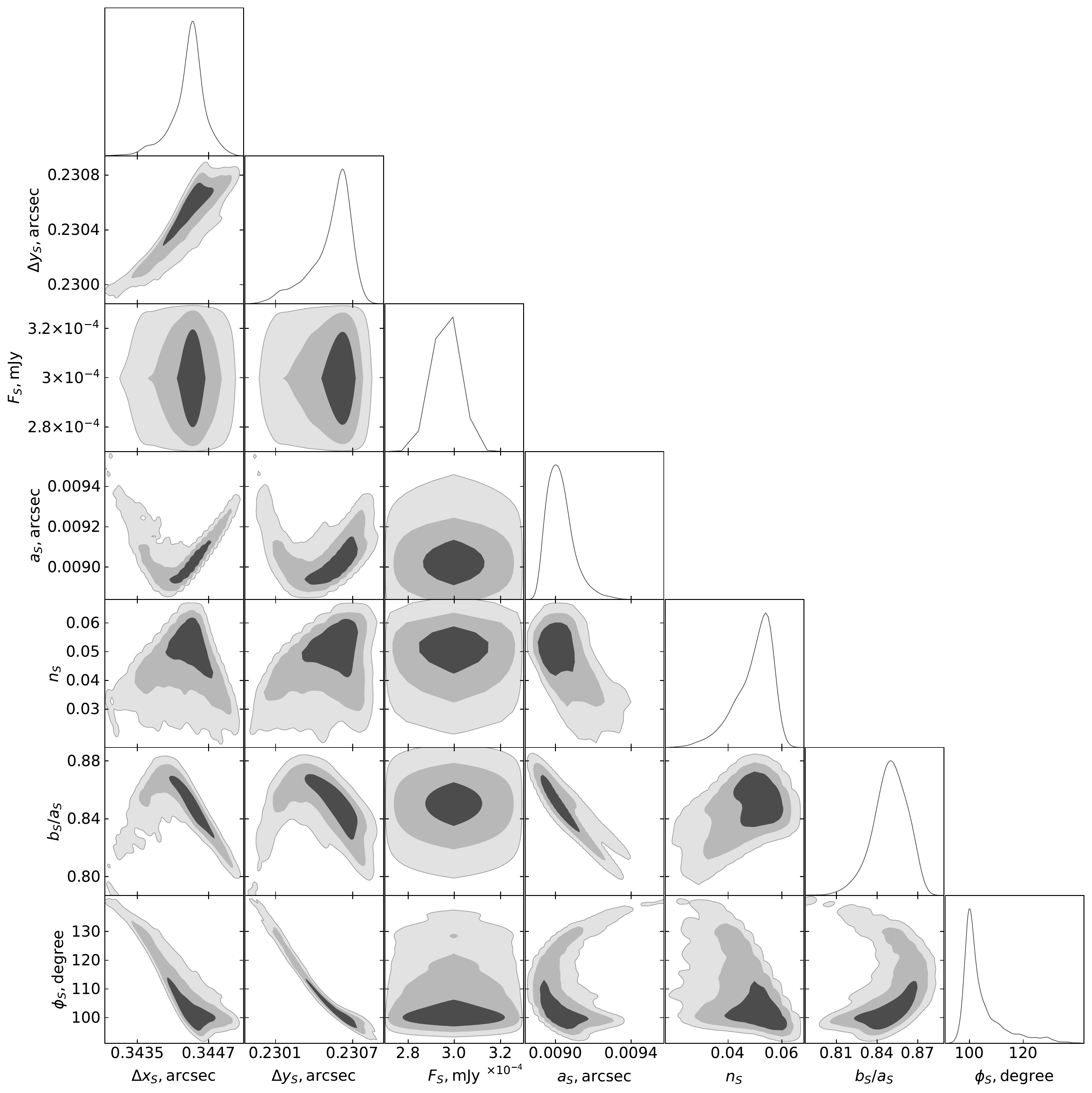}
\caption{Source parameter degeneracy assuming the SIE (${\bar S}_{\nu}$) lens model. The contours show the 68\%. 95\%, and 98\% confidence regions.}
\end{figure*}
\subsection{Source Model}
Secondly, we reconstructed the source by holding the `SIE (${\bar S}_{\nu}$)' lens model fixed and fitting a S{\'e}rsic profile using the software package \texttt{visilens} \citep{hezaveh13a,spilker16}, a Python package for modeling interferometric observations of strong gravitational lensing systems. The source emission was modeled with a S{\'e}rsic profile \citep{sersic63} with seven parameters (columns 2-8 in Table \ref{table:visilensfit}): relative position of the source center to the lens center in the RA and DEC directions $\Delta X_s$ and $\Delta Y_s$, total integrated flux density of the source $F_s$, source major axis $a_s$, S{\'e}rsic profile index $n_s$, source minor to major axis ratio $b_s/a_s$ and source position angle $\phi_s$. The software \texttt{visilens} was modified to make use of the importance nested sampling algorithm \texttt{Multinest} \citep{feroz08,feroz09,feroz19} and its Python interface \texttt{PyMultiNest} \citep{buchner14} for improved fitting efficiency and more precise estimation of the parameter uncertainties for the best-fit model in a multi-model parameter space. The calculation of the log-likelihood function in \texttt{visilens} remains unchanged. A constant efficiency mode of \texttt{MultiNest} with a sampling efficiency of 0.1 and 1000 live points was used. To reduce the number of visibilities to model, the continuum visibility data were averaged over 128 frequency channels in each spectral window spanning 1.875 GHz and 6.048 seconds in time. This is a prudent choice because the response $R_a$ to a point source resulting from this averaging is close to 1 \citep[Equation (6.80) in ][]{thompson17}. In comparison, the source model assuming the `SIE (${\bar S}_{\nu}$)' lens model has higher log-evidence log $Z$ than that assuming the `SIE ({[O {\scriptsize III}]})' lens model in the last column of Table \ref{table:visilensfit}.

The model image from our best lens and source model `SIE (${\bar S}_{\nu}$)' is shown in Figure \ref{modelimage}. All four image positions can be reproduced. Our model predicts an intrinsic source flux density of 0.2963 $\mu$Jy and total magnification of 24.22, corresponding a total flux density of 7.176 mJy after lensing. The model-predicted extended arcs near the triplet images were not observed, due to the low sensitivity of the hybrid array to the angular scales of these arcs discussed in Section \ref{1422:obs}, resulting in a lower total observed flux density of 6.305 mJy. The model predicted S{\'e}rsic source major axis of 9.03 mas corresponds to 66.9 pc at the redshift of this source. This physical size is similar to the narrow-line [O {\scriptsize III}] size of 60 pc \citep{nierenberg14}. The fitted S{\'e}rsic index $n_s$ much smaller than 0.5 describes a sharp cutoff to the source surface profile. This is likely because of the lack of sensitivity to extended emission and the source not resembling the S{\'e}rsic profile.

\subsection{Joint Modeling}
Thirdly, we attempted modeling the visibility data directly by jointly fitting fourteen model parameters for the lens, external shear, and a S{\'e}rsic source using both the original version of \texttt{visilens} and our modified \texttt{visilens} with \texttt{MultiNest} sampler. The best-fit model image produced by the original \texttt{visilens} could not reproduce the relative image positions. The fitted image positions were more widely separated than those in the observation. The poor fits were caused by the low S/N of image D in the visibility data, resulting in large uncertainties in lens mass, lens position and source position. Large uncertainties for the source model parameters are also expected, given the under-constrained lens model and unresolved source. Jointly fitting the lens, external shear, and the source using the modified \texttt{visilens} with \texttt{MultiNest} sampler did not converge to produce a best-fit model. Therefore, we conclude that the joint lens and source model cannot be well constrained by the current data.

The failure of a smooth mass model is not surprising for this source. B1422+231 has been chosen for this study because of its known flux density ratio anomaly. Modeling visibilities directly makes use of both the position and flux density information of the image components. For a system with a flux density ratio anomaly, the best-fit smooth lens model constrained by the image component flux density information is bound to be inaccurate by design. Therefore, lens modeling for systems with flux density ratio anomalies are usually only based on the positions of image components. Previous work has shown that the best-fit SIE+$\gamma_x$ models with and without B1422+231 image flux densities are noticeably different \citep{nierenberg14}. Previous SIE+$\gamma_x$ models for B1422+231 observed in other frequencies can mostly reproduce the image component positions, but they also have difficulties matching the observed component flux densities \citep{kormann94,bradac02,chiba02,nierenberg14,schechter14}. Their best-fit SIE+$\gamma_x$ model parameters can sometimes differ significantly because of the uncertainties in image component positions and flux densities in different frequency bands. The positions of image components are very sensitive to the source and lens positions, in particular, the relative position of the source to the caustics. The S/N of the image components in the dirty image is too low for visually identifying the image component. The \texttt{visilens} software optimizes $\chi^2$ for model visibilities and marginalize over phase errors instead of optimizing the fit to image component positions. The constraints from image component positions cannot be disentangled from the constraints from image component flux densities. This makes modeling visibilities directly for a lens system with anomalous flux density ratios challenging.

\section{Primordial Black Hole Microlensing Simulation} \label{1422:sim}
If a fraction of dark matter is in PBHs, microlensing by a population of PBHs creates a scatter in lensing magnifications of strongly lensed images. We perform PBH microlensing simulations to quantify the effects of PBH dark matter on image component flux density ratios. The brute force inverse ray shooting algorithm, \texttt{GPU-D}, is used to simulate microlensing at the three cusp image components of B1422+231 \citep{thompson10,bate10,vernardos14}. Such ray tracing simulations have traditionally been used to produce magnification maps and light curves for a source lensed by a random field of stars. The total surface mass density $\kappa$ has a smooth component represented by $\kappa_s$ and a compact object componenet $\kappa_c$ consisting of point masses with $\kappa = \kappa_s + \kappa_c$. Following \citet{vernardos14}, we introduce a smooth matter fraction
\begin{equation}
    f_s = \frac{\kappa_s}{\kappa}.
\end{equation}

The three image components of B1422+231, namely, A, B and C, are simulated separately with the same smooth matter fraction $f_s$, but different surface density $\kappa$ and shear $\gamma$ determined by the macromodel for the lens galaxy. Within each simulated area that represents an image component, $\kappa$ and $\gamma$ are assumed to be constant. This assumption is valid because the three unresolved image components have small angular sizes. The variation of $\kappa$ and $\gamma$ within each image component is negligible compared to the uncertainties of the macromodel. For every image component, the differences between the $\kappa$ and $\gamma$ values predicted by different best-fit macromodels in the literature are much larger than their variations within the image component \citep[e.g.][]{mediavilla09,schechter14}. The exact $\kappa$ and $\gamma$ values from the macromodel are not important, because here we are more interested in the scatter of magnification due to PBH microlensing rather than the exact magnification. To also account for magnification uncertainties from the macromodel, PBH microlensing simulations have to be performed for all the macromodels sampled during lens modeling, which is outside the scope of this study. We choose to adopt the ($\kappa$, $\gamma$) values at A (0.38, 0.473), B (0.492, 0.628), and C (0.365, 0.378) from \citet{schechter14} for our simulations, which have been used for stellar microlensing simulations. Our best-fit `SIE (${\bar S}_{\nu}$)' model predicts similar ($\kappa$, $\gamma$) values. We assume that the fraction of matter in dark matter is equal to $\Omega_{DM}/\Omega_{m}=0.842$ \citep{planckxiii16} and that smooth matter consists of non-PBH dark matter and baryonic matter. Our simulations explored the scenarios where the fraction of dark matter in PBHs, $f_{PBH}$, was $10\%$ and $50\%$. The PBH surface density $\kappa_{PBH}$ is given by
\begin{equation}
    \kappa_{PBH} = f_{PBH} \frac{\Omega_{DM}}{\Omega_m}.
\end{equation}
The number of microlenses (i.e. PBHs), $N_{PBH}$, is
\begin{equation}
    N_{PBH} = \frac{\kappa_{PBH} A}{\pi \left<M\right>},
\end{equation}
where $A$ is the area where PBHs are randomly distributed and $\left<M\right>$ is the mean mass of the PBHs following a power law mass function given below with a mass function exponent $\gamma\in(-1, +1)$ \citep{carr17},
\begin{equation}\label{massfunc}
    \psi(M) \propto M^{\alpha} = M^{\gamma-1} \quad (M_{min} < M < M_{max}).
\end{equation}
The mass of a PBH, continuous random variable $M$, satisfying the above mass function can be mapped to a continuous uniform distribution $\mathcal{U}(0,1)$
\begin{equation}
    M \sim \left[(M_{min}^{\alpha+1} - M_{max}^{\alpha+1}) \ \mathcal{U}(0,1) + M_{max}^{\alpha+1} \right]^{\frac{1}{\alpha+1}}.
\end{equation}

\begin{figure}
\epsscale{1.1}
\plotone{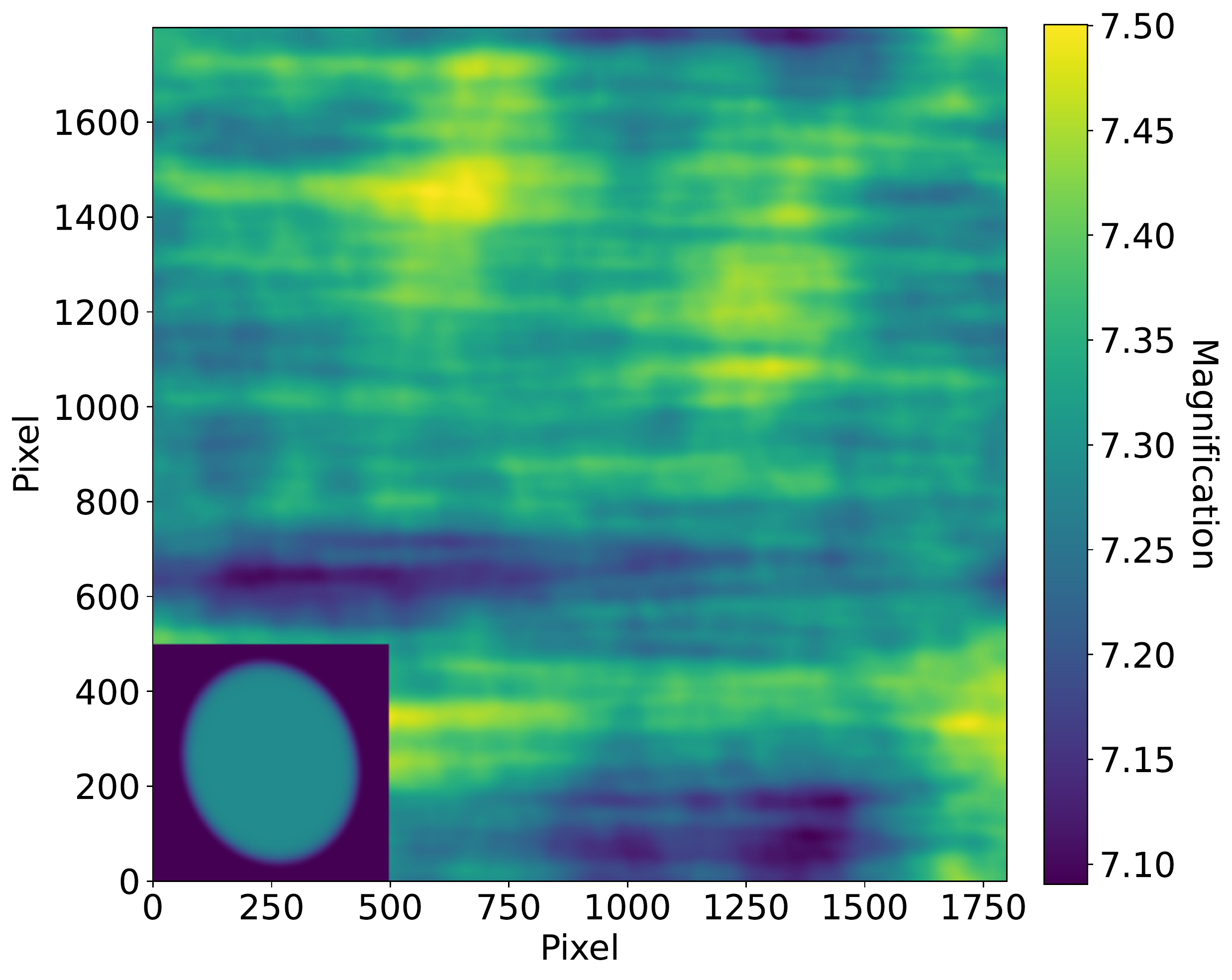}
\caption{A representative portion of the magnification map for image B with $f_{PBH}=0.5$ and $\gamma=-1$ after convolution with our best-fit S{\'e}rsic profile in Table \ref{table:visilensfit}. The shape of the source S{\'e}rsic profile is illustrated in the bottom left corner.}
\label{fig:magmap}
\end{figure}

The two extreme cases where the power law mass function exponent $\gamma=-1$ and $+1$ were explored in our simulations. We generated $N_{PBH}$ random PBH massess and uniformly distributed them in an circular region with area $A$ in the lens plane. This circular region was larger than the area in which light rays were shot. To minimize edge effects, a smaller area within the actual area where light rays were traced was used for estimating the magnification probability distribution. The chosen area must be at least a few times the size of the source in order for the magnification map produced to contain a large enough statistical sample of random source positions on the map. Figure \ref{fig:magmap} shows an example magnification map of image component B of B1422+231 after convolution with our best-fit source S{\'e}rsic profile in Table \ref{table:visilensfit}. Image components with larger surface density have more densely populated PBHs. The large-scale horizontal patterns reflect the shear direction. High resolution magnification maps showing individual caustics are useful for simulating microlensing light curves for a source size comparable to the Einstein radius of the microlenses. However, for the purpose of finding the magnification of a source much larger than the average PBH Einstein radius, the PBH caustics do not need be resolved. Our simulations produced low resolution magnification maps that may contain multiple PBHs per pixel. For each $f_{PBH}$ and $\gamma$ combination, 25 independent 2000$\times$2000 pixels magnification maps were produced. The width of each pixel is equivalent to 3.6 to 26.5 times of the Einstein radius of the average PBH mass, depending on $f_{PBH}$ and $\gamma$. The deflection due to each PBH was still solved exactly. Because it was not necessary to resolve the caustics, the number of light rays needed was drastically decreased, averaging 153.33 light rays per pixel. This significantly reduced the computational cost. To calculate the magnifications of A, B and C at 233 GHz, we convolved the magnification maps with the S{\'e}rsic profile in our best-fit source model in Table \ref{table:visilensfit}, where the source major axis $a_s=9.03$ mas. A smaller source size would allow larger fluctuations in the magnification probability distribution, hence a higher probability for flux density ratio anomaly. Fast Fourier transform implemented in \texttt{Astropy} \citep{astropy13, astropy18} with periodic boundaries that wrap around the magnification map was used for convolution. Varying the source position angle has an negligible effect on the magnification probability distribution.

\begin{figure*}
\epsscale{1.1}
\plottwo{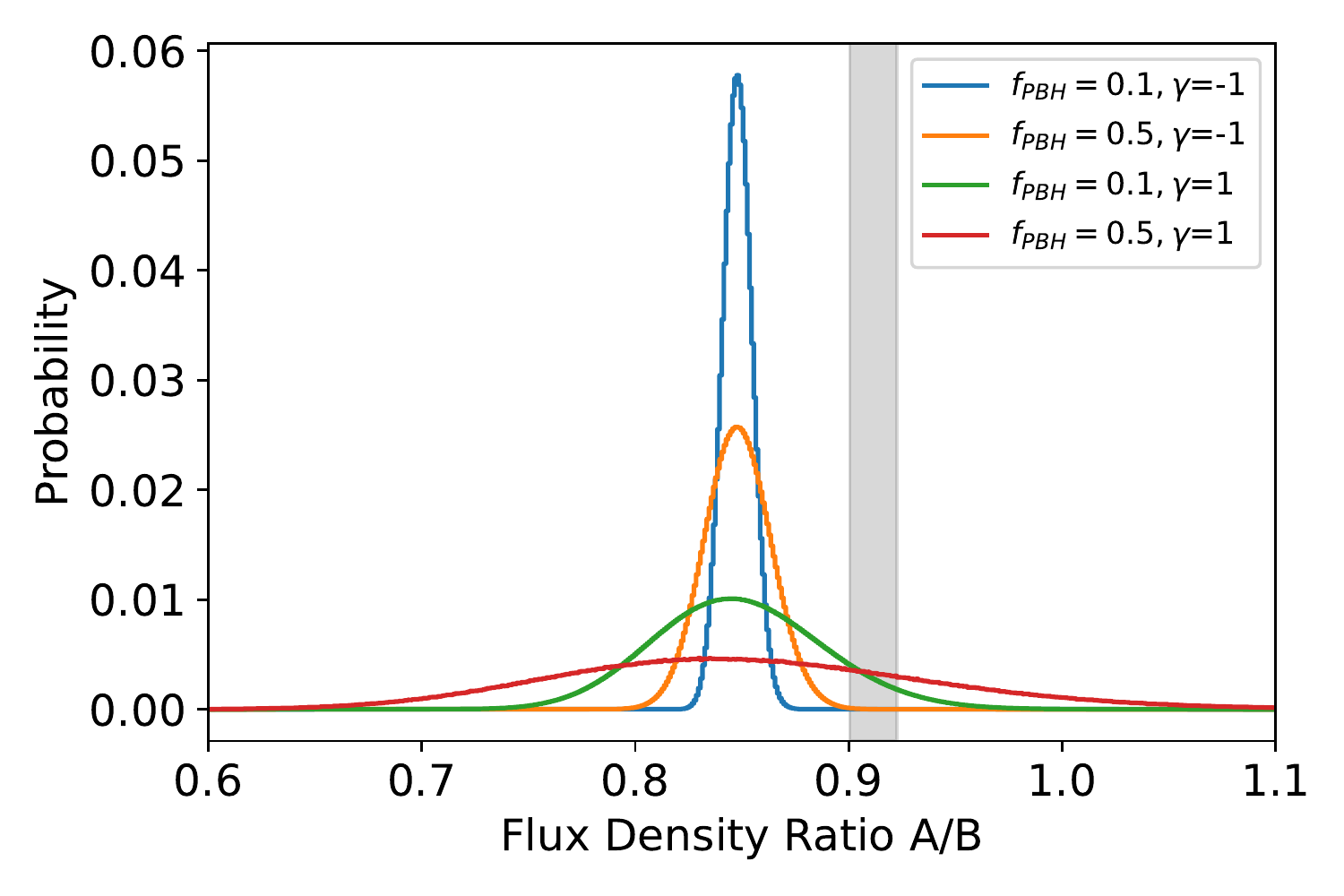}{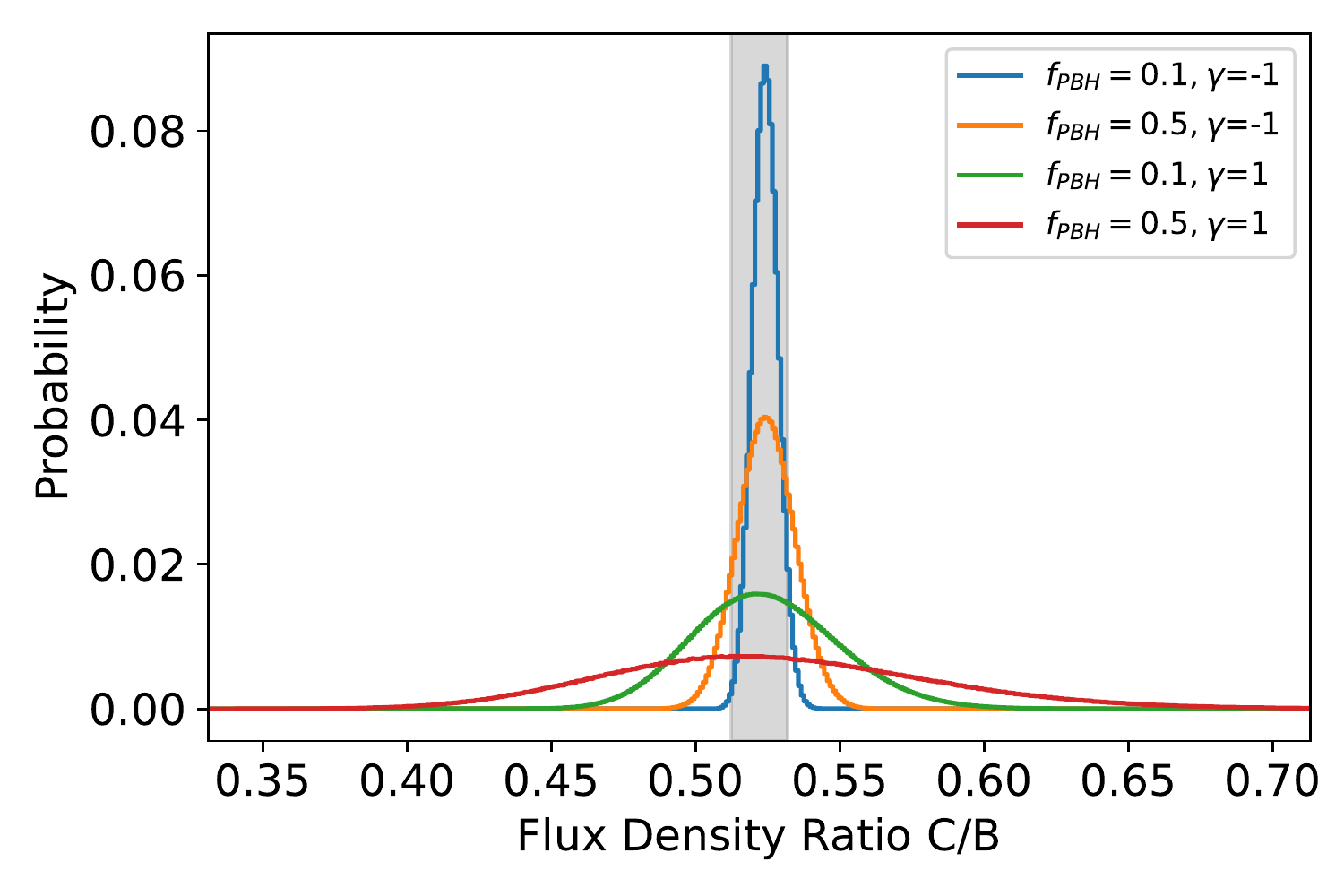}
\plottwo{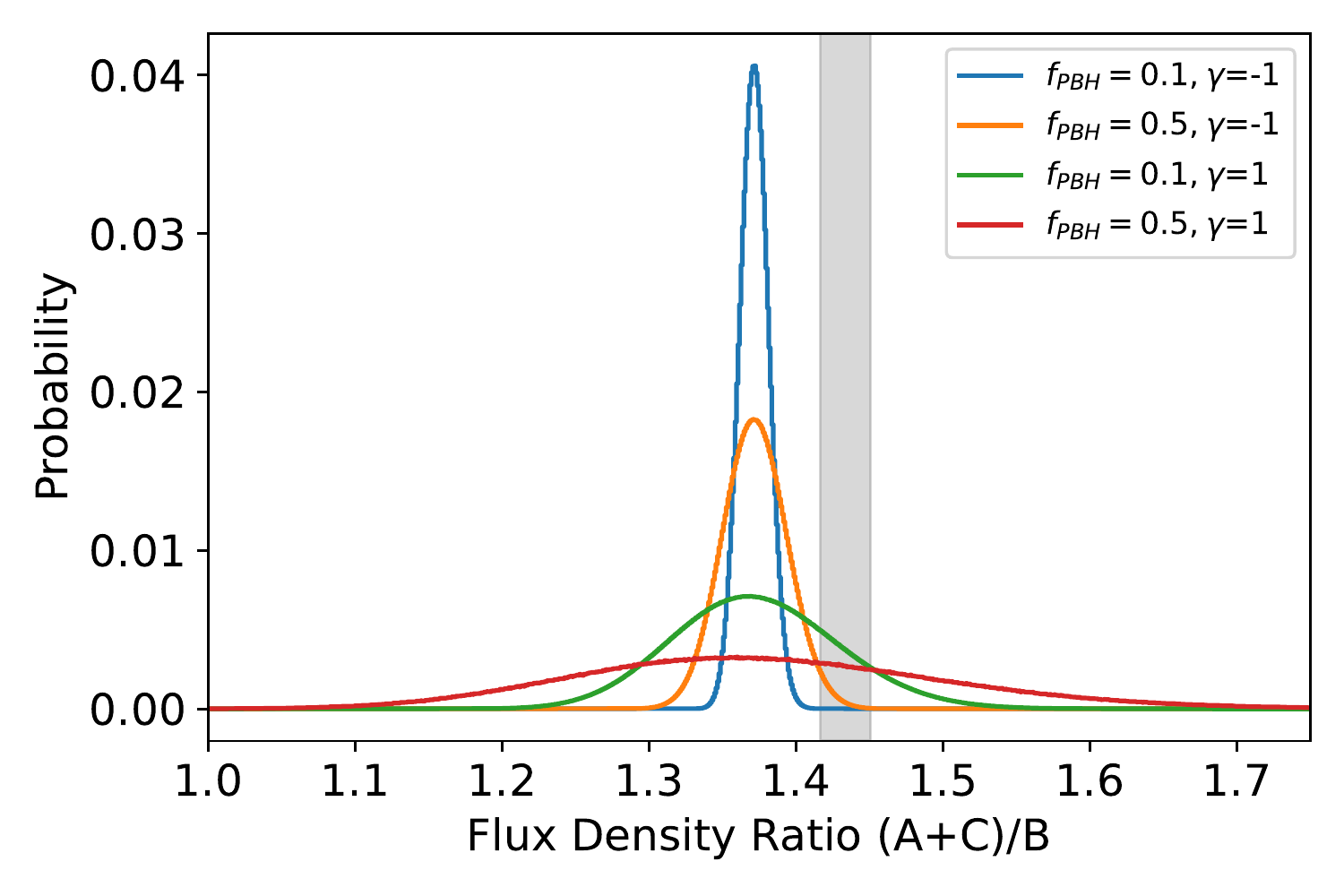}{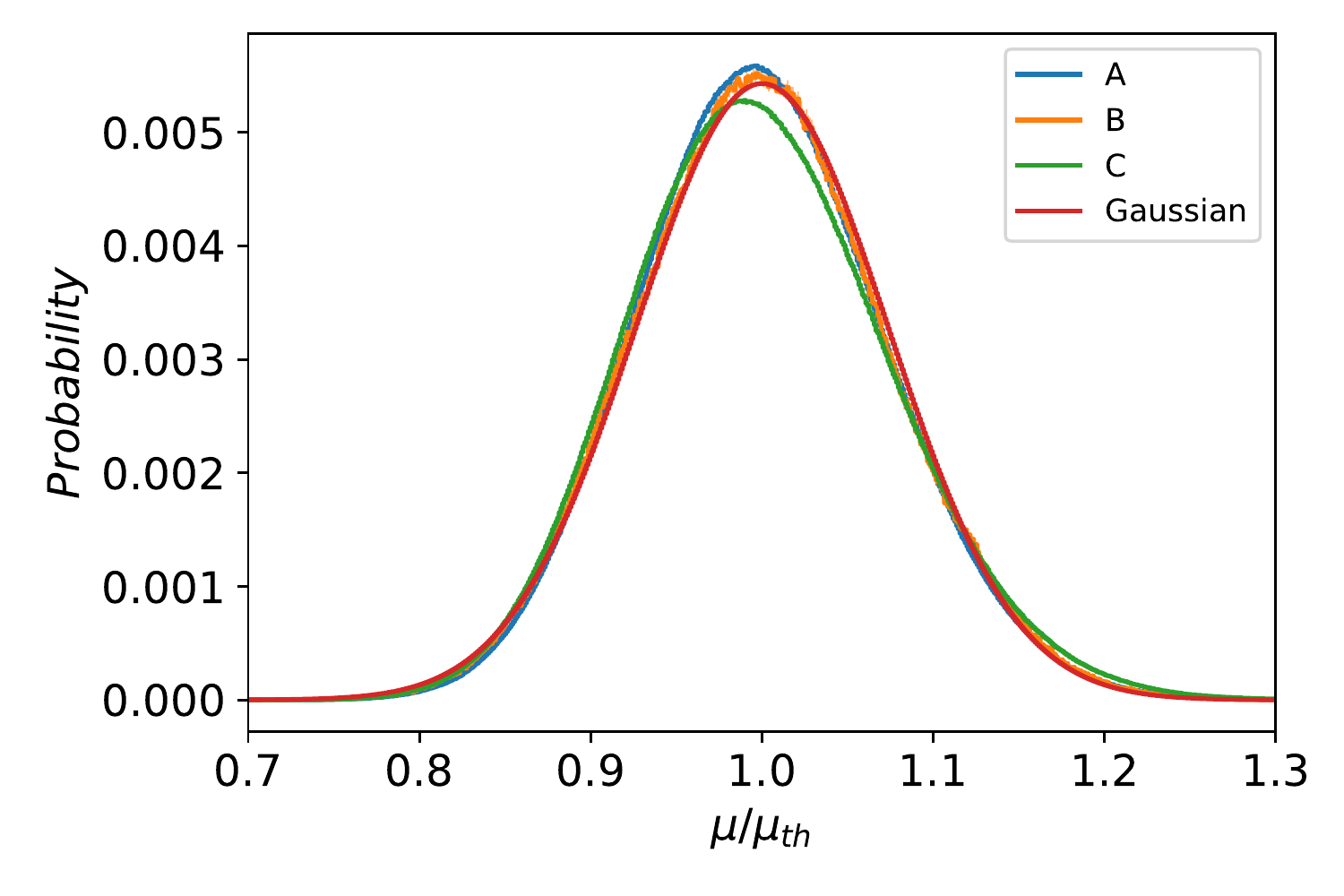}
\caption{Probability distribution function of flux density ratio A/B (upper left), C/B (upper right), and (A+C)/B (lower left) from PBH microlensing simulations, where $f_{PBH}$ is the fraction of dark matter in PBHs. $\gamma$ is the power law index of the PBH mass function as defined in Equation \ref{massfunc}. The grey color bands show our measured flux density ratios within uncertainties from Table \ref{components}. The lower right panel shows the probability distribution functions of image components' magnification when $f_{PBH}=0.5$ and $\gamma=1$, normalized by their respective theoretical magnification predicted by the macromodel. A Gaussian distribution is plotted for comparison. All plotted distributions are histograms with the same bin width 0.001.}
\label{fig:mpd}
\end{figure*}

\begin{deluxetable*}{ccccccccc}
\tablewidth{0pt} 
\tablecaption{P-value and Confidence Interval (CI) of the probability distribution of flux density ratios plotted in Figure \ref{fig:mpd}. The P-values represent the cumulative probabilities of flux density ratio deviating away from the peak of the distribution more than the values measured by ALMA in Table \ref{components}. The confidence interval $z\sigma$ is the proportion of flux density ratios that lie within the smallest symmetric interval ($-z\sigma$, $+z\sigma$) that includes the value measured by ALMA in Table \ref{components}, assuming a Gaussian distribution with standard deviation $\sigma$. \label{table:pvalue}}
\tablehead{
\colhead{Flux Density Ratio} & \multicolumn{4}{c}{$f_{PBH}=0.1$} & \multicolumn{4}{c}{$f_{PBH}=0.5$}\\
\cline{2-9}
\colhead{} & \multicolumn{2}{c}{$\gamma=-1$} & \multicolumn{2}{c}{$\gamma=+1$} & \multicolumn{2}{c}{$\gamma=-1$} & \multicolumn{2}{c}{$\gamma=+1$}\\
\cline{2-9}
\colhead{} & \colhead{P-value} & \colhead{CI} & \colhead{P-value} & \colhead{CI} & \colhead{P-value} & \colhead{CI} & \colhead{P-value} & \colhead{CI}
} 
\startdata 
{  A/B  } & 0 & - & 0.0627 &  1.5322$\sigma$ & 4.373$\times10^{-5}$ & 3.9230$\sigma$ & 0.2414 &    0.7017$\sigma$\\ 
{  C/B  } & 0.3579 & 0.3640$\sigma$ & 0.4829 & 0.04276$\sigma$ & 0.4250 & 0.1891$\sigma$ & 0.4936 & 0.01612$\sigma$\\
{(A+C)/B} & 2.000$\times10^{-8}$ & 5.4909$\sigma$ & 0.1425 & 1.0691$\sigma$ & 0.002989 & 2.7490$\sigma$ & 0.3154 & 0.4806$\sigma$\\
\enddata
\end{deluxetable*}

\section{Simulation Results and Discussion} \label{1422:results}
Figure \ref{fig:mpd} shows the probability distribution function of the flux density ratios and the magnification probability distributions of image component A, B and C as functions of the PBH dark matter fraction, $f_{PBH}$, and the power law index $\gamma$ of the PBH mass function. We find that for our assumed source major axis of 9.03 mas, the probability distribution function of the flux density ratio (A+C)/B and the magnification probability distributions of the three cusp image components are all well-described by Gaussian distributions. This is in contrast to the asymmetric multi-modal distribution functions of magnification for small source sizes \citep{schechter02,schechter04,schechter14}. Because the assumed source size at 233 GHz is much larger than the Einstein radius of an average PBH, the large fluctuations of magnification probability distribution as the source crosses the caustics are essentially removed when convoluted with the source. Nonetheless, the widths of the probability distribution functions shown in Figure \ref{fig:mpd} indicate that there is a still non-negligible scatter in magnifications and flux density ratios caused by the microlensing of PBH dark matter. Table \ref{table:pvalue} lists the P-values and confidence intervals of flux density ratio found by our simulations. 

We find that a larger fraction of PBH dark matter, $f_{PBH}$, or a larger power index $\gamma$ for the PBH mass function both noticeably widens the probability distribution function of the flux density ratios and the magnification probability distributions. With a power law index $\gamma=-1$, the number of PBHs sharply declines as a function of PBH mass. With a power law index $\gamma=1$, the number of PBHs is approximately constant in each mass bin in the specified mass range $10-1000M_{\odot}$. The difference in the power law index $\gamma$ most directly affects the average PBH mass. The number density of PBHs is proportional to the fraction of PBH dark matter. The lower right panel of Figure \ref{fig:mpd} shows that for given $f_{PBH}$ and $\gamma$, the probability distributions of magnifications normalized by macromodel predicted magnification ($\mu/\mu_{th}$) are very close to Gaussian distribution and they does not depend on the surface mass density or shear. More clumpy dark matter, either due to a higher fraction of PBH dark matter or fewer but more massive PBHs, increases the probability of the flux density ratio anomaly. The scatters of flux density ratios and magnifications in Figure \ref{fig:mpd} are conservative, because stellar populations are considered part of the smooth matter and any scatter produced by stellar microlensing is ignored. Figure \ref{fig:mpd} shows that a change in the PBH mass function alters the probability of having a large flux density ratio anomaly more significantly than a change of PBH dark matter fraction from 10\% to 50\%.  Without the inclusion of a massive DMS, PBH dark matter between $10-1000M_{\odot}$ with a flat mass function alone can produce the observed flux density ratio anomaly. This highlights the need to include the effects of PBH dark matter when considering the causes of flux density ratio anomalies of multiply imaged strong lenses.

Assuming PBH dark matter is the only cause for the flux density ratio anomaly of B1422+231 in our ALMA observation, our simulations show that the probability of observing a flux density ratio (A+C)/B$>$1.434 is 31.54\% if 50\% dark matter is $10-1000M_{\odot}$ PBHs with a flat mass function (power law index $\gamma=+1$), and 14.25\% if 10\% dark matter is $10-1000M_{\odot}$ PBHs with a flat mass function (see Table \ref{table:pvalue}). The cases where the PBH mass function is highly skewed towards low masses (power law index $\gamma=-1$) have negligibly small probabilities, irregardless of the PBH dark matter fraction. \citet{carr17} claims that any power law index$\gamma$ outside $(-1,+1)$ is not permitted. The logarithmic PBH mass function proposed by \citet{carr17} also highly skews towards low mass PBHs, so the probabilities assuming a logarithmic PBH mass function should be similarly small compared to those from the power law when $\gamma=-1$. Hence, if PBH dark matter has a mass function highly skewed towards low masses, its contribution to the flux density ratio anomaly should be extremely small even if a significant fraction of dark matter is PBHs. Recent constraints on $f_{PBH}$ allow 100\% dark matter being PBHs in the intermediate mass range $10M_{\odot}<M<10^3M_{\odot}$ \citep{carr20}. However, more updated constraints suggest that the fraction of PBH dark matter in this mass range is very small, while PBHs below $10^{-10} M_{\odot}$ may still make up the majority of dark matter \citep{carr21b}. Such low mass PBHs can be treated as smooth matter microlensing simulations. Given that the P-values are small for $f_{PBH}$, it is very unlikely that PBH dark matter is the only cause for our measured anomalous flux density ratio if $f_{PBH}<0.1$.

The P-values are expected to be higher if there is PBH dark matter outside the $10-1000M_{\odot}$ mass range. An additional fraction of PBH dark matter from other mass ranges will result in more microlensing events and widen the probability distributions of the flux density ratios. Treating stars as microlenses instead of a smooth mass component will have a similar effect. In our simulations, stars in the mass range of $10-1000M_{\odot}$ are part of the smooth baryonic matter surface density. Therefore, our statistics in Table \ref{table:pvalue} are conservative estimates of the effects of PBH dark matter on flux density ratios.

DMSs are likely a major contributor to the flux density ratio anomaly of B1422+231 according to previous studies \citep{chiba02,nierenberg14,xu15}. Their models add a DMS modeled by a singular isothermal sphere to the macrolens model near image A. It is possible that the primary cause for the flux density ratio anomaly in B1422+231 is a DMS and the secondary cause is PBH dark matter. To more conclusively disentangle of the effects of PBH dark matter from DMS, more comprehensive mass models for multi-frequency observations, including low mass perturbers in a wide mass range such as DMs, IMBHs and stars, need to be compared. One key difference between DMS and PBH in causing flux density ratio anomalies is that PBHs as a fraction of dark matter affect all the image components, whereas usually one DMS affects one image component at a time. In principle, there are many low-mass DMSs, but their number density is still much lower than that of PBHs. One can potentially distinguish DMS-caused flux density ratio anomaly from PBHs by checking if only one of the image components has a anomalous flux density.

The biggest uncertainty in constraining the fraction of dark matter in PBHs is the unknown PBH mass function. The uncertainties from the smooth mass model and any massive DMS are the main source of systematic errors of the flux density ratio probability distribution functions. Incorporating the results of an ensemble of microlensing simulations with varied PBH mass functions into a full Bayesian analysis of a lens model including a smooth halo component, DMSs, and PBHs has the potential to provide constraints on PBH dark matter fraction and mass function. Quadruply-lensed compact mm bright quasars with larger flux density ratio anomalies than B1422+231 will give tighter constraints on PBH dark matter. 

\section{Conclusions} \label{1422:conclusions}
The mass distribution in the foreground lens galaxy of multiply imaged strong gravitational lenses are often not well described by simple smooth lens models. Quadruple lenses such as B1422+231 show flux density ratio anomalies in multi-frequency observations. In radio wave bands, the flux density ratio anomaly is caused by DMSs. We argue that PBH dark matter can be the cause of flux density ratio anomalies in the mm-wave band. In addition to DMSs, PBH dark matter may also be a secondary contributor to the flux density ratio anomaly. In optical and X-ray bands, stellar microlensing is the primary cause for flux density ratio anomaly, but both DMSs and PBH dark matter should be taken into account for lens modeling.

We present the first continuum imaging of B1422+231 using ALMA at 233 GHz. Four unresolved point sources are detected with a total continuum flux density of $6.305\pm0.082$ mJy. The detected compact emission is consistent with the spectral energy distribution of synchrotron emission from the quasar's AGN with almost no thermal dust emission. The flux density ratio (A+B)/C is measured to be $1.434\pm0.017$, consistent with the measurements in radio and mid-infrared frequencies. The smooth lens model is not well constrained by the visibility data directly. Our lens models from mm-wave image plane modeling are similar to those from narrow-line [O {\scriptsize III}] \citep{nierenberg14}. Assuming a S{\'e}rsic source profile, the source major axis is estimated to be $9.03^{+0.15}_{-0.12}$ mas, corresponding to 66.9 pc at the redshift of this source, similar to the intrinsic [O {\scriptsize III}] size \citep{nierenberg14}.

Our ray tracing microlensing simulations of the three cusp image components show that PBHs in the intermediate mass range $10M_{\odot}<M<10^3M_{\odot}$ as a fraction of dark matter can produce flux density ratio anomalies. The magnification probability distribution of image components are well described by Gaussian distributions. A larger fraction of dark matter in PBHs and a less negative power law index for the PBH mass function can both increase the probability of flux density ratio anomaly. Assuming PBHs are the only cause of flux density ratio anomaly for B1422+231, we determine an upper limit to the probability of (A+C)/B$>$1.434 to be 31.54\% for up to 50\% dark matter being PBHs, and 14.25\% for up to 10\% dark matter being PBHs. Our measured flux density ratio (A+B)/C=1.434 is within 0.4806$\sigma$ confidence interval of the prediction for 50\% dark matter in PBHs with a flat mass function. PBHs with a highly skewed mass function towards low masses has very low probability of being the only cause for the observed flux density ratio anomaly. This is the first study that quantifies the effects of PBHs on the flux density ratio anomaly in mm-wave bands. Our analysis places new constraints on the fraction of PBH dark matter and the PBH mass function using a quadruple lens with flux density ratio anomaly and ray tracing simulations.

\begin{acknowledgments}
We thank John P. McKean for insightful discussions about the interpretation of the ALMA observation and Garrett K. Keating for sharing the flux density measurement of B1422+231 from the Submillimeter Array. This paper makes use of the following ALMA data: ADS/JAO.ALMA\#2018.1.00915.S. ALMA is a partnership of ESO (representing its member states), NSF (USA) and NINS (Japan), together with NRC (Canada), MOST and ASIAA (Taiwan), and KASI (Republic of Korea), in cooperation with the Republic of Chile. The Joint ALMA Observatory is operated by ESO, AUI/NRAO and NAOJ. Support for this work was provided by the NSF through award SOSPA6-019 from the NRAO. The National Radio Astronomy Observatory is a facility of the National Science Foundation operated under cooperative agreement by Associated Universities, Inc. This research is part of the Blue Waters sustained-petascale computing project, which is supported by the National Science Foundation (awards OCI-0725070 and ACI-1238993) the State of Illinois, and as of December, 2019, the National Geospatial-Intelligence Agency. Blue Waters is a joint effort of the University of Illinois at Urbana-Champaign and its National Center for Supercomputing Applications. DW acknowledges support from the Netherlands Organization for Scientific Research (NWO) (Project No. 629.001.023) and the Chinese Academy of Sciences (CAS) (Project No. 114A11KYSB20170054). DW thanks the Center for Information Technology of the University of Groningen for their support and for providing access to the Peregrine high performance computing cluster.
\end{acknowledgments}

\vspace{5mm}
\facility{ALMA}

\software{CASA \citep{mcmullin07}, Astropy \citep{astropy13, astropy18}, visilens \citep{hezaveh13a,spilker16}, MultiNest \citep{feroz09,feroz19}, PyMultiNest \citep{buchner14}, GPU-D \citep{thompson10,bate10,vernardos14}, Numpy \citep{vanderwalt11}, glafic\citep{oguri10}, GetDist\citep{lewis19}.}

\bibliography{B1422_ALMA}{}
\bibliographystyle{aasjournal}

\end{document}